\documentclass[aps,preprint,preprintnumbers,amsmath,amssymb]{revtex4}
\usepackage{amsmath,mathrsfs,amsbsy,color,graphicx,bm,amsthm,amsfonts}
\usepackage{units}
\usepackage{bbm}
\usepackage{times}
\usepackage{dcolumn}
\usepackage{mathrsfs}
\usepackage{amsmath,amssymb,epsfig}

\begin{document}

\title{Quantum steering for continuous variable in de Sitter space }
\author{  Cuihong Wen$^{1}$, Jieci Wang$^{1}$\footnote{Email: jcwang@hunnu.edu.cn}, and Jiliang Jing$^{1}$\footnote{Email: jljing@hunnu.edu.cn}}
\affiliation{$^1$ Department of Physics, and Collaborative Innovation Center for Quantum Effects \\
and Applications,
 Hunan Normal University, Changsha, Hunan 410081, China
}


\begin{abstract}
We study the distribution of quantum steerability  for continuous variables between two causally disconnected open charts in de Sitter space.   It is shown that  quantum steerability  suffers from  ``sudden death" in de Sitter space, which is quite different from the behaviors of   entanglement and discord because the latter always survives and  the former vanishes only in the limit of infinite curvature. It is found  that  the attainment of maximal steerability asymmetry indicates a transition between unidirectional  steerable and bidirectional steerable. Unlike in the flat space, the asymmetry of quantum steerability can be completely  destroyed in the limit of infinite curvature  for the conformal and massless scalar fields  in  de Sitter space.

\end{abstract}

\vspace*{0.5cm}

\maketitle
\section{Introduction}

Einstein-Podolsky-Rosen steering \cite{epr}  captures the fact that one observer, can nonlocally  manipulates, or steers, the state of  the other subsystem by performing  measurements on one-half of an entangled state.  Originally realized by Schr\"odinger \cite{schr,schr2},  quantum steerability
has been understood as a form of intermediate nonlocal correlation between Bell nonlocality and entanglement. It is recognized  that  steerability is an important   resource for quite a few  quantum information processing tasks, such as one-sided device-independent quantum key distribution \cite{steering1,steering11} and subchannel discrimination \cite{steering12}. In practice, different from the Bell tests, the demonstration of quantum steerability which is  free of detection and locality loopholes is in reach  \cite{Kocsis}, which make  quantum steerability a ponderable
 concept in quantum information theory.  For the foregoing reasons, quantum steerability has recently attracted increasing interest  both from theoretical \cite{wiseman,Skrzypczyk,Walborn, Bowles,steering2, steering3, steering4, steering5, wang2016, jieci1} and experimental \cite{reid,Saunders,Handchen,Sun,zeng} perspectives.
Most recently, Tischler {\it et.al} reported   an experimental demonstration of unidirectional  steering, which is free either of  the  restrictions on the type of allowed measurements or of assumptions about the quantum state at hand \cite{Tischler}.

On the other hand, it is of great interest to study the behavior of quantum entanglement  and other quantum correlations for discrete variables between causally disconnected regions  in the context of cosmology \cite{Ball:2005xa,Fuentes:2010dt,Nambu:2011ae, Kanno16, wang2015,Kanno151,Kanno152,Kanno153,Choudhury}. In particular,  we know that any two mutually separated regions eventually become causally disconnected in the exponentially expanding de Sitter space.  The Bogoliubov transformations
between  the open chart vacua and the  Bunch-Davies vacuum  which  have support on both regions of a free massive scalar field were derived in~\cite{Sasaki:1994yt}. Later, quantum correlations of scalar fields~\cite{Maldacena:2012xp,Albrecht18, Kanno:2014lma,Iizuka:2014rua}, Dirac fields~\cite{Kanno:2016qcc} and axion fields ~\cite{Choudhury:2017bou,Choudhury2019} were studied in de Sitter space.
In~\cite{Kanno:2014ifa, Dimitrakopoulos:2015yva}, the authors studied the observable effect of quantum information on the cosmic microwave background  since there are some entanglement between causally separated regions in de Sitter space.   Like entanglement,  quantum steerability  is one kind of   nonlocal quantum correlation which admits no equivalent in classical physics. Therefore, it is important to study the quantum steering between  causally disconnected regions of de Sitter space  because one can never communicate classically between them.

In this paper we present a quantitative investigation on the distribution of   steerability  in de Sitter space  by employing  an operational measure of  quantum steerability for continuous variable systems  \cite{Adesso2015} .
 We consider the sharing of quantum steerability among three subsystems:  subsystems $A$ (observed by Alice) and  $B$ (observed by Bob) in  region $R$, and subsystem $\bar B$ observed by an imaginary observer anti-Bob who is restricted to region $L$ of de Sitter space. We will  derive the  phase-space description of quantum state evolution  for continuous variables  basing on the Bogoliubov transformation
between  the open chart vacua and the  Bunch-Davies vacuum.   It is found that the quantum steerability  between Alice and Bob is apparently affected by the  curvature of de Sitter space when the mass parameter $\nu$ approaches to the limit of    $\nu=1/2 $ (conformal) and $\nu=3/2$ (massless). At the same time, Bob and antiBob can steer each other  when the curvature is strong enough  even though they are separated by the event horizon,  which verifies the nonlocal peculiarity of quantum steerability.

The outline of the paper is as follows. In Sec. II we discuss the dynamics of mode functions and Bogoliubov transformations  in de Sitter space.  In Sec. III we  review the definition and measure of bipartite Gaussian quantum steerability. In Sec. IV we study the distribution of Gaussian quantum steerability in de Sitter space. The last section is devoted to a brief summary.

\section{Quantization of  scalar field in de Sitter space \label{model}}

We consider a free scalar field $\phi$ with mass $m$ initially observed by two experimenters, Alice and Bob, in tthe Bunch-Davies vacuum of de Sitter space.   The coordinate frames of open charts in de Sitter space  can be obtained by analytic continuation from the Euclidean metric. As shown in Fig. (1), the spacetime geometry of  de Sitter space in open charts  is divided into three parts denoting by $R$, $L$ and $C$, respectively. We assume that the observer  Bob is restricted to region $R$, which is causally disconnected from region $L$.  The metrics for the two causally  disconnected open charts $R$ and $L$ in the de Sitter space
are given by
\begin{eqnarray}
ds^2_R&=&H^{-2}\left[-dt^2_R+\sinh^2t_R\left(dr^2_R+\sinh^2r_R\,d\Omega^2\right)
\right]\,,\nonumber\\
ds^2_L&=&H^{-2}\left[-dt^2_L+\sinh^2t_L\left(dr^2_L+\sinh^2r_L\,d\Omega^2\right)
\right]\,,
\end{eqnarray}
where $d\Omega^2$ is the metric on the two-sphere and $H^{-1}$ is the Hubble radius.

\begin{figure}
\includegraphics[scale=0.27]{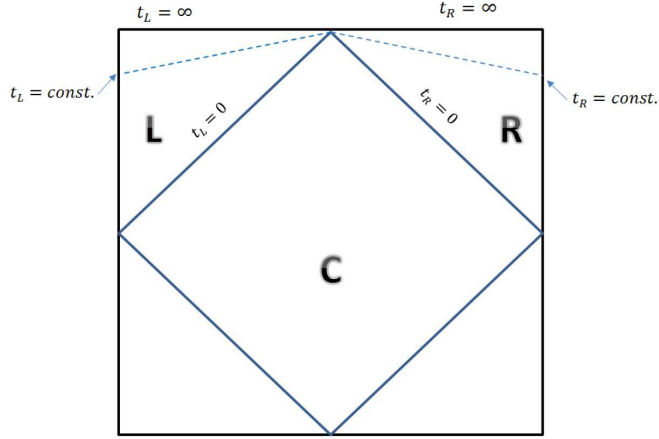}
\caption{(Color online) The Penrose diagram of the de Sitter space, in which
$L$ and $R$ are the two causally disconnected regions
described by the open charts. }\label{Fig1}
\end{figure}

Solving the Klein-Gordon equation in different regions, one obtains
\begin{eqnarray}\label{solutions1}
u_{\sigma p\ell m}(t_{R(L)},r_{R(L)},\Omega)&\sim&\frac{H}{\sinh t_{R(L)}}\,
\chi_{p,\sigma}(t_{R(L)})\,Y_{p\ell m} (r_{R(L)},\Omega)\,,\qquad \nonumber\\
-{\rm\bf L^2}Y_{p\ell m}&=&\left(1+p^2\right)Y_{p\ell m}\,,
\end{eqnarray}
with  $Y_{p\ell m}$ being harmonic functions on the three-dimensional hyperbolic space. In Eq. (\ref{solutions1}) $\chi_{p,\sigma}(t_{R(L)})$ are positive frequency mode functions supporting  on the $R$ and $L$ regions ~\cite{Sasaki:1994yt}
\begin{eqnarray}
\chi_{p,\sigma}(t_{R(L)})=\left\{
\begin{array}{l}
\frac{e^{\pi p}-i\sigma e^{-i\pi\nu}}{\Gamma(\nu+ip+\frac{1}{2})}P_{\nu-\frac{1}{2}}^{ip}(\cosh t_R)
-\frac{e^{-\pi p}-i\sigma e^{-i\pi\nu}}{\Gamma(\nu-ip+\frac{1}{2})}P_{\nu-\frac{1}{2}}^{-ip}(\cosh t_R)
\,,\\
\\
\frac{\sigma e^{\pi p}-i\,e^{-i\pi\nu}}{\Gamma(\nu+ip+\frac{1}{2})}P_{\nu-\frac{1}{2}}^{ip}(\cosh t_L)
-\frac{\sigma e^{-\pi p}-i\,e^{-i\pi\nu}}{\Gamma(\nu-ip+\frac{1}{2})}P_{\nu-\frac{1}{2}}^{-ip}(\cosh t_L)
\,,
\label{solutions}
\end{array}
\right.
\end{eqnarray}
where $P^{\pm ip}_{\nu-\frac{1}{2}}$ are the associated Legendre functions
and  $\sigma=\pm 1$  distinguishing  the 
independent solutions for each region. These solutions can be normalized by the  factor
$
N_{p}=\frac{4\sinh\pi p\,\sqrt{\cosh\pi p-\sigma\sin\pi\nu}}{\sqrt{\pi}\,|\Gamma(\nu+ip+\frac{1}{2})|}\,.
$ In addition, $p$ is a positive real parameter normalized by $H$,  and $\nu$ is a mass parameter
$
\nu=\sqrt{\frac{9}{4}-\frac{m^2}{H^2}}\,
$.
Note that the effect of the curvature of the three-dimensional hyperbolic space starts to appear around $p\sim 1$ \cite{Kanno16, Albrecht18}. In addition, the effect of  curvature gets stronger as $p$ becomes smaller than $1$. Therefore, $p$ can be regarded as the curvature parameter of the de Sitter space.
It is known that there are two special values for the mass parameter  $\nu$:
 $\nu=1/2$ ($m^2=2H^2$) for the conformally coupled
massless scalar field, and $\nu=3/2$ for the minimally coupled massless limit.

The scalar field can be
 expanded in terms of the creation and annihilation operators:
\begin{eqnarray}
\hat\phi(t,r,\Omega)
=\frac{H}{\sinh t}\int dp \sum_{\ell,m}\phi_{p\ell m}(t)Y_{p\ell m}(r,\Omega)
\,,
\end{eqnarray}
where the Fourier mode field operator $
\phi_{p\ell m}(t)\equiv
\sum_\sigma\left[\,a_{\sigma p\ell m}\,\chi_{p,\sigma}(t)
+a_{\sigma p\ell -m}^\dagger\,\chi^*_{p,\sigma}(t)\right]$ has been
 introduced, and
$a_{\sigma p\ell m}|0\rangle_{\rm BD}=0$ is the  annihilation operator of the Bunch-Davies vacuum.
For simplicity, hereafter we omit the indices $p$, $\ell$, $m$ of $\phi_{p\ell m}$,
$a_{\sigma p\ell m}$ and $a_{\sigma p\ell -m}^\dag$.
Similarly, the mode functions and the associated Legendre
functions  are rewritten in  simple forms:
 $\chi_{p,\sigma}(t)\rightarrow\chi^{\sigma}$,
$P_{\nu-1/2}^{ip}(\cosh t_{R,L})\rightarrow P^{R, L}$, and
$P_{\nu-1/2}^{-ip}(\cosh t_{R,L})\rightarrow P^{R*, L*}$.

Then we consider the positive frequency
mode functions
\begin{eqnarray}
\varphi^q=\left\{
\begin{array}{ll}
\tilde{N}_p^{-1}P^q~&\mbox{in region}~q\,,
\\
0~ &\mbox{in the opposite region}\,,
\end{array}
\right.
\quad\tilde{N}_p=\frac{\sqrt{2p}}{|\Gamma(1+ip)|}\,,
\label{varphi}
\end{eqnarray}
for the $R$ or $L$ vacuum which
 are defined only on the $q=(R, L)$ region, respectively. Since the Fourier mode field operator should
be the same under the change of mode functions, we can relate the operators $(a_\sigma,a_\sigma^\dag)$ and $(b_q,b_q^\dag)$
 by a Bogoliubov transformation
\begin{eqnarray}
\phi(t)=a_\sigma\,\chi^\sigma+a_\sigma^\dag\,\chi^\sigma{}^*
=b_q\,\varphi^q+b_q^\dag\,\varphi^q{}^*\,,
\label{fo}
\end{eqnarray}
where  the creation and annihilation operators ($b_q,b_q^\dag$) in different regions are introduced to ensure
 $b_q|0\rangle_{q}=0$.

Using the Bogoliubov transformation between the operators, the Bunch-Davies vacuum can be constructed from the vacuum states
over $|0\rangle_{q}$ in regions $R$ and $L$, which is
\begin{eqnarray}
|0\rangle_{\rm BD}\propto\exp\left(\frac{1}{2}\sum_{i,j=R,L}
m_{ij}\,b_i^\dagger\, b_j^\dagger\right) |0\rangle_R|0\rangle_L\,,
\label{bogoliubov1}
\end{eqnarray}
where $m_{ij}$ is a symmetric matrix determined by
 $a_\sigma|0\rangle_{\rm BD}=0$:
\begin{eqnarray}\nonumber
m_{ij}=\frac{\sqrt{2}\,e^{-p\pi}}{\sqrt{\cosh 2\pi p+\cos 2\pi\nu}}
\left(
\begin{array}{cc}
\cos \pi\nu & i\sinh p\pi \vspace{1mm}\\
i\sinh p\pi & \cos \pi\nu \\
\end{array}
\right)\,.
\label{mij}
\end{eqnarray}
We can see that the Bunch-Davies vacuum
is in fact an entangled two mode squeezed state in the ${\cal H}_R\otimes{\cal H}_L$ Hilbert space.
 It is worth noting  that the density matrix $\rho=|0\rangle_{\rm BD}\,{}_{\rm BD}\langle0|$ is diagonal only
for $\nu= 1/2$ or $3/2$.

To make the calculation easier for tracing out the degrees of freedom in the $L$ space later, a diagonal  density matrix like
\begin{eqnarray}
|0\rangle_{\rm BD} = N_{\gamma_p}^{-1}
\exp\left(\gamma_p\,c_R^\dagger\,c_L^\dagger\,\right)|0\rangle_{R'}|0\rangle_{L'}\,,
\label{bogoliubov3}
\end{eqnarray}
is required. To this end
we  introduce new operators $c_q=(c_R,c_L)$ that satisfy  \cite{Kanno16, Albrecht18}
\begin{eqnarray}
c_R = u\,b_R + v\,b_R^\dagger \,,\qquad
c_L = u^*\,b_L + v^*\,b_L^\dagger\,.
\label{bc}
\end{eqnarray}
Apparently, this Bogoliubov transformation does
not mix the operators in the open chart $R$ and those in open chart $L$. Note that the condition $|u|^2-|v|^2=1$ is assumed
to ensure the commutation relation
$[c_i,(c_j)^\dagger]=\delta_{ij}$.
The normalization factor $N_{\gamma_p}$ in Eq. (\ref{bogoliubov3}) is given by
\begin{eqnarray}
N_{\gamma_p}^2
=\left|\exp\left(\gamma_p\,c_R^\dagger\,c_L^\dagger\,\right)|0\rangle_{R'}|0\rangle_{L'}
\right|^2
=\frac{1}{1-|\gamma_p|^2}\,.
\label{norm2}
\end{eqnarray}
Considering the definition  of $c_R$ and $c_L$ in Eq. (\ref{bc})  and the consistency relations from Eq. (\ref{bogoliubov3}), it is  demanded that
 $c_R|0\rangle_{\rm BD}=\gamma_p\,c_L^\dag|0\rangle_{\rm BD}$,
$c_L|0\rangle_{\rm BD}=\gamma_p\,c_R^\dag|0\rangle_{\rm BD}$, which give  \cite{Kanno16, Albrecht18}
\begin{eqnarray}
\nonumber\gamma_p=\frac{1}{2\zeta}
\left[-\omega^2+\zeta^2+1-\sqrt{\left(\omega^2-\zeta^2-1\right)^2-4\zeta^2}\,\right]\,,
\label{gammap}
\end{eqnarray}
where we defined $\omega\equiv m_{RR} = m_{LL}$ and $\zeta\equiv m_{RL}=m_{LR}$ in Eq.~(\ref{mij}). Putting the  matrix elements of  Eq.~(\ref{mij})
into Eq.~(\ref{gammap}), we obtain
\begin{eqnarray}
\gamma_p = i\frac{\sqrt{2}}{\sqrt{\cosh 2\pi p + \cos 2\pi \nu}
 + \sqrt{\cosh 2\pi p + \cos 2\pi \nu +2 }}\,.
\label{gammap2}
\end{eqnarray}
For the conformally coupled
massless scalar ($\nu=1/2$) and the minimally coupled massless scalar ($\nu=3/2$),  $\gamma_p$ simplifies to $|\gamma_p|=e^{-\pi p}$.

\section{Measurement of quantum steerability for continuous variables \label{GSteering}}

In this section we introduce the definition and  measurement of  quantum steerability for  continuous variables. We consider a bosonic  bipartite continuous variable quantum system \cite{weedbrook} represented
by $(n+m)$  modes. The bipartite state  consist of two subsystems: the first  subsystem is observed by  Alice ($A$) with $n$ modes and the second  subsystem  for Bob ($B$) of $m$ modes.  For each mode $i$,  the corresponding phase-space operators $\hat x_i^{A(B)},\,\,\hat p_i^{A(B)}$ are defined by $\hat a_i^A=\frac{\hat x_i^A+i\hat p_i^A}{\sqrt{2}}$ and $\hat a_i^B=\frac{\hat x_i^B+i\hat p_i^B}{\sqrt{2}}$.  These phase-space variables  grouped for convenience into the vector
$\hat R = (\hat x_1^A,\hat p_1^A, \ldots ,\hat x_n^A,\hat p_n^A,\hat x_1^B,\hat p_1^B, \ldots ,\hat x_m^B,\hat p_m^B)^{\sf T}$, satisfying the canonical commutation relations $[{{{\hat R}_i},{{\hat R}_j}} ] = i{\Omega _{ij}}$, with $\Omega  =  \bigoplus_1^{n+m} {{\ 0\ \ 1}\choose{-1\ 0}}$ being the symplectic form. The character of a Gaussian state ${\rho _{AB}}$  is
 is completely prescribed by its first and second statistical moments. The latter is a covariance matrix  with elements ${\sigma _{ij}} = \text{Tr}\big[ {{{\{ {{{\hat R}_i},{{\hat R}_j}} \}}_ + }\ {\rho _{AB}}} \big]$ and can always be
put into a block form
\begin{equation}\label{CM}
\sigma_{AB} = \left( {\begin{array}{*{20}{c}}
   A & C  \\
   {{C^{\sf T}}} & B  \\
\end{array}} \right).
\end{equation} Here the submatrices $A$ and $B$ are the covariance matrixes corresponding to the reduced states of Alice's and Bob's subsystems respectively. In addition,  a covariance matrix $\sigma_{AB}$  that can describe a physical quantum state if and only if  (\textit{iff} ) the \textit{bona fide} uncertainty principle relation
\begin{equation}\label{bonafide}
{\sigma _{AB}} + i\,({\Omega _{AB}} ) \ge 0,
\end{equation}
is  satisfied.

Now let us give the definition of quantum  steerability in continuous variable systems. We consider a pair of local observables $R_A$ ( on $A$ with outcome $r_A$) and $R_B$  (on $B$ with outcome $r_B$) in a bipartite state $\rho_{AB}$. After Alice performs a set of measurements $\mathcal{M}_A$, the state is  steerable  \textit{iff} it is \textit{not} possible
 to express the joint probability as \cite{wiseman}
\begin{equation}\label{steering} P\left( {{r_A},{r_B}|{R_A},{R_B},{\rho _{AB}}} \right) = \sum\limits_\lambda  {{\wp_\lambda }} \, \wp\left( {{r_A}|{R_A},\lambda } \right)P\left( {{r_B}|{R_B},{\rho _\lambda }} \right), \end{equation}
where  ${\wp_\lambda }$ and $\wp \left( {{r_A}|{R_A},\lambda }\right)$  are
probability distributions, involving
the local hidden variable $\lambda$. In addition,  $P\left( {{r_B}|{R_B},{\rho _\lambda }} \right)$  is the conditional  probability distribution associated to the extra condition of being evaluated on the state $\rho_\lambda$. Like the Bell nonlocality, quantum steering  is exhibited in a state
\textit{iff}  the correlations between A and B cannot be explained
by a local hidden variable model. In other words, at least one measurement set is required to violate the expression when  ${\wp_\lambda }$ is fixed across all measurements.

As  proposed in \cite{wiseman}, a Gaussian state $\rho_{AB}$ is  $A\to B$ steerable  \textit{iff} the condition
\begin{equation}\label{nonsteerm}
\nonumber{\sigma _{AB}} + i\,({0_A} \oplus {\Omega _B}) \ge 0,
\end{equation}
is violated by Alice's Gaussian measurements. 
Employing Eq. (\ref{CM}), we can see that the  inequality given in  Eq. (\ref{nonsteer}) equals to two simultaneous conditions: (i) $A > 0$, and (ii) ${M^B_{\sigma}} + i{\Omega _B} \ge 0$, with $M^B_{\sigma} = B - {C^{\sf T}}{A^{ - 1}}C$ being the Schur complement of $A$ in the CM $\sigma_{AB}$. Note that
the first  condition is always verified because $A$ is a physical  covariance matrix. Therefore, $\sigma_{AB}$ is $A \to B$ steerable \textit{iff} the symmetric and positive definite $2m \times 2m$  matrix $M^B_{\sigma}$ is not a {\it bona fide} covariance matrix \cite{wiseman}.

According to Williamson's theorem \cite{williamson}, the symmetric matrix $M^B_{\sigma}$ is diagonalized by a symplectic transformation $S_B$  such that $S_B M^B_{\sigma} S_B^{\sf T}=\text{diag}\{\bar{\nu}^B_1,\bar {\nu}^B_1,\ldots,\bar{\nu}^B_m,\bar {\nu}^B_m\}$, where $\{\bar{\nu}^B_{j}\}$ are the symplectic eigenvalues of $M^B_{\sigma}$.
Then  the degree of  $A \to B$ steerability can be measured  by  \cite{Adesso2015}
\begin{equation}\label{GSABm}
{\cal G}^{A \to B}(\sigma_{AB}):=
\max\bigg\{0,\,-\sum_{j:\bar{\nu}^B_j<1} \ln(\bar{\nu}^B_j)\bigg\}\,,
\end{equation}
which quantifies the amount by which the condition given by Eq. (\ref{nonsteerm}) fails to be fulfilled.
This is the  Gaussian $A \to B$ steerability, which is invariant under local symplectic operations, it vanishes \textit{iff} the state described by  Eq. (\ref{CM})  is nonsteerable by Gaussian measurements. In other words, the  $A \to B$ steerability in fact quantifies the degree by which the condition (\ref{nonsteerm}) fails to be fulfilled by Alice's measurement. 

If  the steered party Bob has one mode only, the   $A \to B$ steerability acquires this form \cite{Adesso2015}
\begin{eqnarray}\label{GSAB}
 {\cal G}^{A \to B}(\sigma_{AB}) =
 \max\big\{0,\, {\cal S}(\sigma_A) - {\cal S}(\sigma_{AB})\big\}\,, \label{GS1}
\end{eqnarray}
with ${\cal S}(\sigma) = \frac12 \ln( \det \sigma)$ being  the R\'enyi-$2$ entropy  \cite{renyi}.
Also, the Gaussian $B \to A$ steerability can be defined by swapping the roles of $A$ and $B$ in Eq. (\ref{GS1}). In a quantum information scenario, quantum steerability  corresponds to the  task of quantum information distribution by an untrusted party \cite{wiseman}. If Alice and Bob share a  $A \to B$ steerable state, the  untrusted Alice is able to convince Bob  that the shared state is entangled by performing local measurements and classical communication \cite{wiseman}.

\section{Distribution of Gaussian quantum steerability in de Sitter space \label{tools}}
\subsection{Reduction of  quantum steerability between initially correlated modes }

We assume that Alice is a  global observer who stays at the Bunch-Davies vacuum, while Bob is an open chart observer resides  in the R region of the de Sitter space.
 The initial state of the modes is prepared by an entangled Gaussian two-mode squeezed state in  the Bunch-Davies vacuum, which is described by  the covariance matrix
\begin{eqnarray}\label{inAR}
\sigma^{\rm (G)}_{AB}(s)= \left(\!\!\begin{array}{cccc}
\cosh (2s)&0&\sinh (2s)&0\\
0&\cosh (2s)&0&-\sinh (2s)\\
\sinh (2s)&0&\cosh (2s)&0\\
0&-\sinh (2s)&0&\cosh (2s)
\end{array}\!\!\right),
\end{eqnarray}
where  $s$ is the squeezing of the  initial state. The derivation of the covariance matrix for a  two-mode squeezed state is given in Appendix A.    In Eq. (\ref{inAR}), the basic vectors of the covariance matrix are  $|ij\rangle=|i\rangle_{A}|j\rangle_{B}$, which denote the state are observer by Alice (A) and Bob (B). Considering there is no initial correlation between the entire state $\sigma^{\rm (G)}_{AB}(s)$ and the subsystem observed by anti-Bob, the initial covariance matrix of the entire system is   $\sigma^{\rm (G)}_{AB}(s) \oplus I_{\bar B}$.

As showed in \cite{Kanno16},  the Bunch-Davies vacuum for a  global  observer can be expressed as a two-mode squeezed state of the $R$ and $L$ vacua
\begin{eqnarray}
\nonumber|0\rangle_{\rm BD}=\sqrt{1-|\gamma_B|^2}\,\sum_{n=0}^\infty\gamma_B^n|n\rangle_L|n\rangle_R\,,
\label{bogoliubov2}
\end{eqnarray}
where $\gamma_B$ is the  squeezing parameter given in Eq. (\ref{gammap2}). In the Fock space, the  two-mode squeezed state can be obtained by
 $|0\rangle_{\rm BD}=\hat{U}_{R,L}(\gamma_p)|0\rangle_{R}|0\rangle_{L}$, where
 $\hat{U}_{R,L}(\gamma_p)=e^{\gamma_p(\hat{c}^\dagger_{\text{R}}\hat{c}^\dagger_{\text{L}}-
\hat{c}_{\text{R}}\hat{c}_{\text{L}})}$  is the two mode squeezing operator. In the phase space, such transformation  can be expressed by a symplectic  operator
\begin{eqnarray}\label{cmtwomode}
 S_{B,\bar B}(\gamma_B)= \frac{1}{\sqrt{1-|\gamma_B|^2}}\left(\!\!\begin{array}{cccc}
1&0&|\gamma_B|&0\\
0&1&0&-|\gamma_B|\\
|\gamma_B|&0&1&0\\
0&-|\gamma_B|&0&1
\end{array}\!\!\right)
\end{eqnarray}
where   $|kl\rangle=|k\rangle_{B}|l\rangle_{\bar B}$, which denotes that squeezing transformation are performed to the bipartite state shared between Bob and anti-Bob ($\bar B$).

Under this transformation, the mode observed by Bob is mapped into two open charts. That is to say,
an extra set of modes $\bar{B}$  becomes relevant from the perspective of a  observer in the open charts. Therefore, a complete description of the system involves three modes, mode $A$ described by Alice, mode $B$ described by the Bob in the $R$ chart, and mode $\bar{B}$  by a  hypothetical observer anti-Bob confined in the $L$ chart. The covariance matrix of the  entire state is given by \cite{adesso3}
\begin{eqnarray}\label{in34}
\nonumber\sigma_{AB \bar B}(s,\gamma_B) &=& \big[I_A \oplus  S_{B,\bar B}(\gamma_B)\big] \big[\sigma^{\rm (G)}_{AB}(s) \oplus I_{\bar B}\big]\\&& \big[I_A \oplus  S_{B,\bar B}(\gamma_B)\big]\,,
\end{eqnarray} 
where $S_{B,\bar B}(\gamma_B)$ is the phase-space representation of the two-mode squeezing transformation given in Eq. (\ref{cmtwomode}). For detail please see Appendix B.

Because Bob in  chart $R$ have no access to the modes in the causally disconnected $L$ region, we must therefore trace over the inaccessible modes.  Taking the
trace over mode $\bar B$ in chart $L$,  one obtains
covariance matrix $\sigma_{AB}(s,\gamma_B)$ for Alice and Bob
\begin{eqnarray}\label{CM1}
\sigma_{AB}(s,\gamma_B)= \left(\!\!\begin{array}{cccc}
\cosh(2s)&0&\frac{\sinh(2s)}{\sqrt{1-|\gamma_B|^2}}&0\\
0&\cosh(2s)&0&-\frac{\sinh(2s)}{\sqrt{1-|\gamma_B|^2}}\\
\frac{\sinh(2s)}{\sqrt{1-|\gamma_B|^2}}&0&\frac{|\gamma_B|^2+\cosh(2s)}{1-|\gamma_B|^2}&0\\
0&-\frac{\sinh(2s)}{\sqrt{1-|\gamma_B|^2}}&0&\frac{|\gamma_B|^2+\cosh(2s)}{1-|\gamma_B|^2}
\end{array}\!\!\right).
\end{eqnarray}
Employing Eq. (\ref{GSAB}),   the $A \to B$ Gaussian  steerability is found to be
 \begin{eqnarray}
 {\cal G}^{A \to B}=
\mbox{$\max\big\{0,\,  \ln {\frac{\cosh(2s)(1-|\gamma_B|^2)}{1 +|\gamma_B|^2 \cosh(2 s) }}\big\}$}. \label{GSab}
\end{eqnarray}

From Eq. (\ref{GSab}) we can see that the $A \to B$ Gaussian  steerability depends not only the squeezing parameter $s$, but also the curvature parameter $p$ and mass parameter $\nu$ of the de Sitter space. To check if the  quantum steerability is  symmetric in  de Sitter space, we also  calculate  the steerability $ {\cal G}^{B \to A}$
 \begin{eqnarray}
 {\cal G}^{B \to A} =
\mbox{$\max\big\{0,\,  \ln {\frac{\cosh(2s)+|\gamma_B|^2}{1 +|\gamma_B|^2 \cosh(2 s) }}\big\}$}.\label{GSba}
\end{eqnarray}
Since the $A\to B$ steering and the $B\to A$ steering are defined in terms of  measurements performed by different observers,
the quantum steering is asymmetry in general. In this paper the asymmetry between A and B appears because the squeezing transformation only acts between Bob and anti-Bob. As we can see from Eq. (\ref{in34}), the symplectic operator for the entire system is  $\big[I_A \oplus  S_{B,\bar B}(\gamma_B)\big]$, which leads the asymmetry between Alice and Bob.

\begin{figure}[htbp]
\centering
\includegraphics[height=2.0in,width=2.5in]{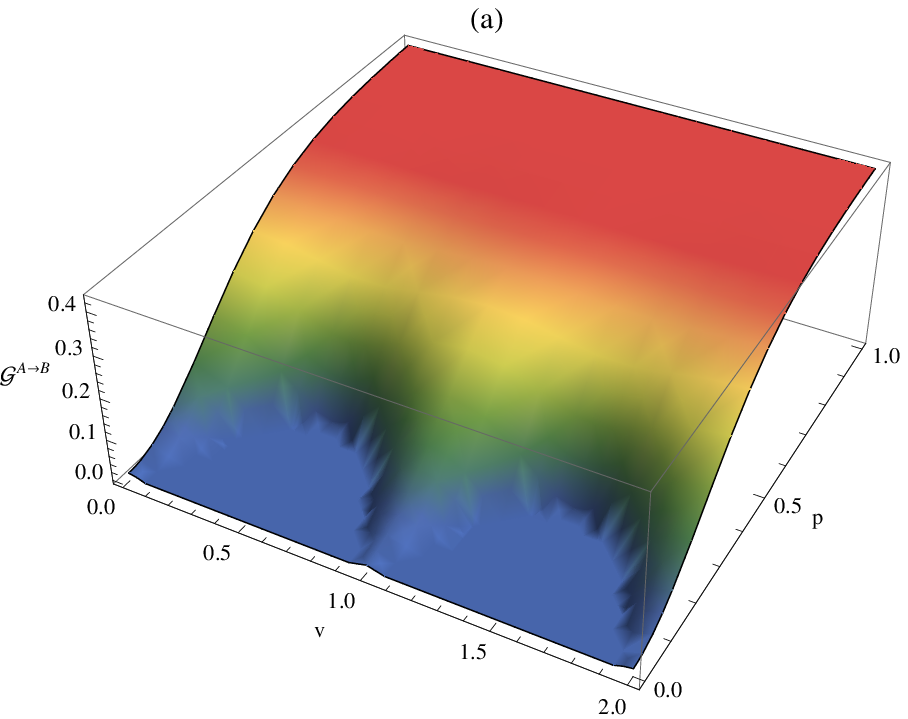}
\includegraphics[height=2.0in,width=2.5in]{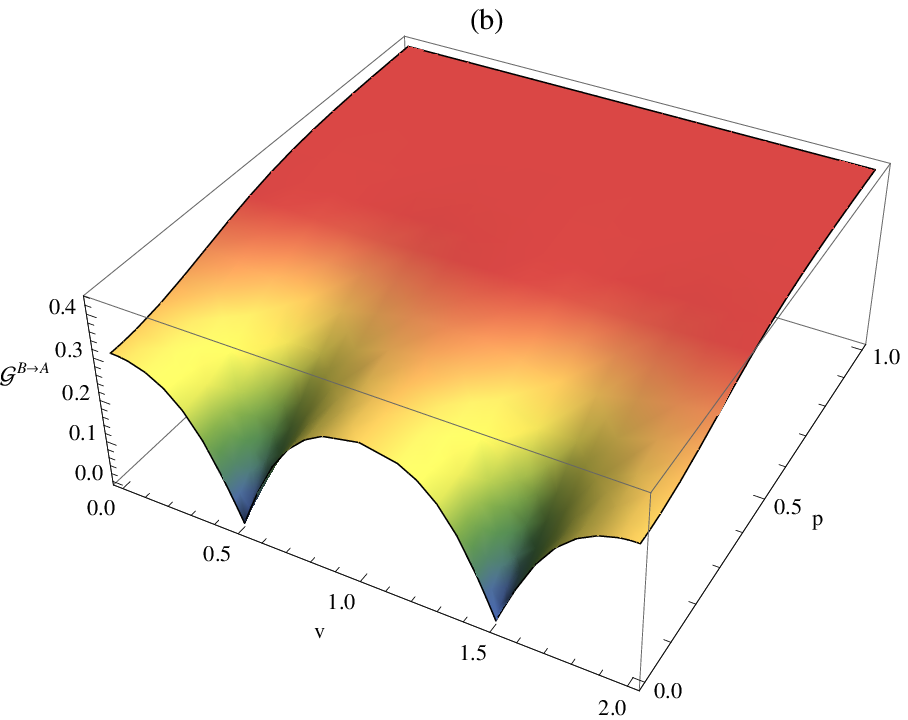}
\caption{ (Color online). The Gaussian quantum steerability $ {\cal G}^{A \to B}$ (left) and $ {\cal G}^{B \to A}$ (right)  as functions of   the curvature parameter $p$ and mass parameter $\nu$ of the de Sitter space. The initial  squeezing parameter of the two-mode squeezed state is fixed as $s=0.5$.}\label{Fig2}
\end{figure}

In Fig. (1) we plot the steerability   $ {\cal G}^{A \to B}$ (left) and $ {\cal G}^{B \to A}$ (right)  as functions of   the curvature parameter $p$ and mass parameter $\nu$ for a fixed squeezing $s=0.5$. We can see that both  the $A \to B$ and  $B \to A$ steerability monotonically decrease with  the decrease of  curvature parameter $p$, which means that space curvature in de Sitter space will destroy the steerability between the initially  modes. However, the quantum steerability is   apparently affected by the  curvature of de Sitter space only around $\nu=1/2 $ (conformal scalar limit) and $\nu=3/2$ (massless scalar limit).  It is shown that for  $\nu=1/2$ (conformal) and $\nu=3/2$ (massless), the $B \to A$ steerability vanishes only in the limit of infinite curvature  $p\to 0$.  However,  the $A \to B$ steerability  suffers from  ``sudden death" when the parameters satisfy  $|\gamma_B|=\sqrt{\frac{1-sech(2s)}{2}}$, which is quite different from the behavior  entanglement and discord  in de Sitter space. It was found that quantum discord always survives \cite{Kanno16} while   entanglement  negativity vanishes only in the limit of infinite curvature  \cite{Maldacena:2012xp,Albrecht18, Kanno:2014lma,Iizuka:2014rua}.

The ``sudden death" of  quantum steerability  indicates the fact that it reduces to zero for  finite curvature  in de Sitter space, while entanglement  vanishes only in the limit of infinite curvature. That is to say, comparing with entanglement, quantum steerability  is more sensitive under the influence of space  curvature. The physical interpretation of this behavior is very intuitive.   It is known that  a quantum state with $\eta_0 \leq \eta < \eta_e$ is  entangled, where  $\eta_0=\mu_A \mu_B + |\mu_A - \mu_B|$,  $\eta_e=\sqrt{\mu_A^2+\mu_B^2-\mu_A^2\mu_B^2}$ and $\mu_{A(B)} = 1/\sqrt{\det A (B)}$ \cite{Adesso2015}. However, a quantum state which satisfies  $\eta \geq \{\mu_A, \mu_B\}$is nonsteerable; a state with $\eta < \mu_B$ is $A \to B$ steerable; while a state with $\eta < \mu_A$ is $B \to A$ steerable, which are within the entangled region.This again proves the fact that quantum steerability is an intermediate nonlocal correlation between  entanglement and  Bell nonlocality . 

\subsection{Generating  quantum steerability between initially uncorrelated modes }

 To explore the distribution of quantum steerability in de Sitter space, we have to know the behavior of steerability between all the bipartite pairs in the tripartite quantum system.
Tracing over the modes in $B$, we obtain the covariance matrix
$\sigma_{A\bar B}(s,\gamma_B)$ between the mode observed by Alice in the $R$ region  and anti-Bob in $L$ region
\begin{eqnarray}\label{CM2}
\sigma_{A\bar B}(s,\gamma_B)= \left(\!\!\begin{array}{cccc}
\cosh(2s)&0&\frac{|\gamma_B| \sinh(2s)}{\sqrt{1-|\gamma_B|^2}}&0\\
0&\cosh(2s)&0&\frac{|\gamma_B| \sinh(2s)}{\sqrt{1-|\gamma_B|^2}}\\
\frac{|\gamma_B|\sinh(2s)}{\sqrt{1-|\gamma_B|^2}}&0&\frac{1+|\gamma_B|^2\cosh(2s)}{1-|\gamma_B|^2}&0\\
0&\frac{|\gamma_B|\sinh(2s)}{\sqrt{1-|\gamma_B|^2}}&0&\frac{1+|\gamma_B|^2\cosh(2s)}{1-|\gamma_B|^2}
\end{array}\!\!\right).
\end{eqnarray}
Interestingly,  we find the mode described by Alice the mode described by anti-Rob  {\it cannot} steer each other  because $ {\cal G}^{A \to \bar B} = {\cal G}^{\bar B \to A} =0$ for any  parameters. In fact, this bipartite state is separable under the  Peres-Horodecki separability criterion for continuous variable systems \cite{Simon}.

 We also interested in the steerability between mode $B$ in the $R$ region  and anti-Bob in $L$ region, which are  separated by the event horizon of the de Sitter space.
Tracing over the modes in $A$, we obtain the covariance matrix
$\sigma_{B\bar B}(s,\gamma_B)$ for Bob and anti-Bob

\begin{eqnarray}\label{CM22}
\sigma_{B\bar B}(s,\gamma_B)= \left(\!\!\begin{array}{cccc}
\frac{|\gamma_B|^2+\cosh(2s)}{1-|\gamma_B|^2}&0&\frac{2|\gamma_B|\cosh^2(s)}{1-|\gamma_B|^2}&0\\
0&\frac{|\gamma_B|^2+\cosh(2s)}{1-|\gamma_B|^2}&0&-\frac{2|\gamma_B|\cosh^2(s)}{1-|\gamma_B|^2}\\
\frac{2|\gamma_B|\cosh^2(s)}{1-|\gamma_B|^2}&0&\frac{1+|\gamma_B|^2\cosh(2s)}{1-|\gamma_B|^2}&0\\
0&-\frac{2|\gamma_B|\cosh^2(s)}{1-|\gamma_B|^2}&0&\frac{1+|\gamma_B|^2\cosh(2s)}{1-|\gamma_B|^2}
\end{array}\!\!\right).
\end{eqnarray}

Then we calculate the $B \to \bar B$ and $\bar B \to B$  steerability, which are found to be
 \begin{eqnarray}
 {\cal G}^{B \to \bar B} =
\mbox{$\max\big\{0,\,  \ln {\frac{1+sech(2s)|\gamma_B|^2}{1 -|\gamma_B|^2  }}\big\}$},\label{GSb-ba}
\end{eqnarray}
and
 \begin{eqnarray}
 {\cal G}^{\bar B \to B} =
\mbox{$\max\big\{0,\,  \ln {\frac{sech(2s)+|\gamma_B|^2}{1 -|\gamma_B|^2  }}\big\}$}, \label{GSba-b}
\end{eqnarray}
respectively.

\begin{figure}[htbp]
\centering
\includegraphics[height=2.0in,width=2.6in]{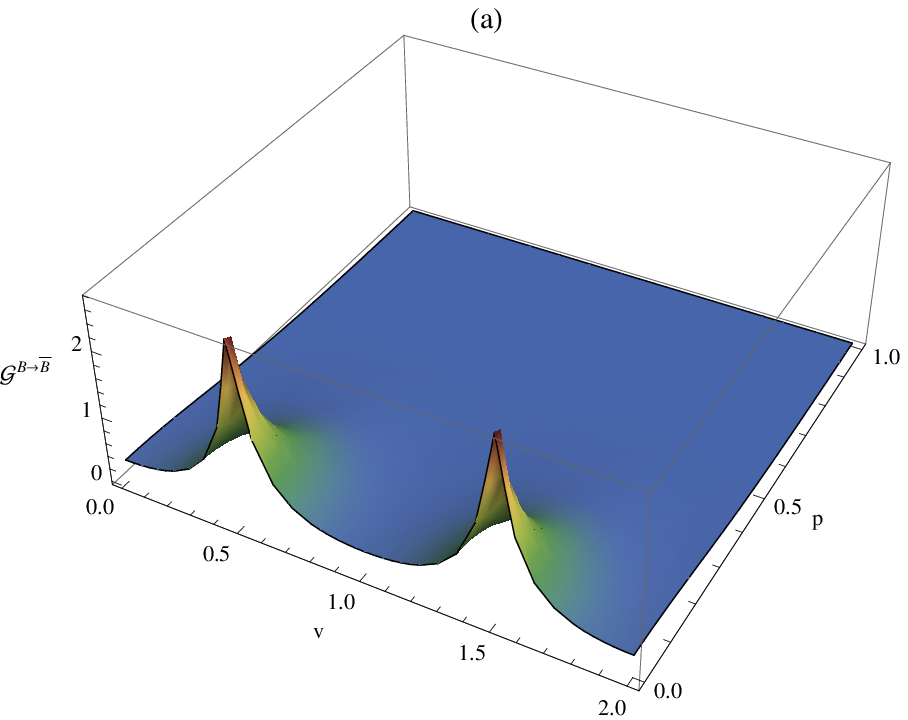}
\includegraphics[height=2.0in,width=2.6in]{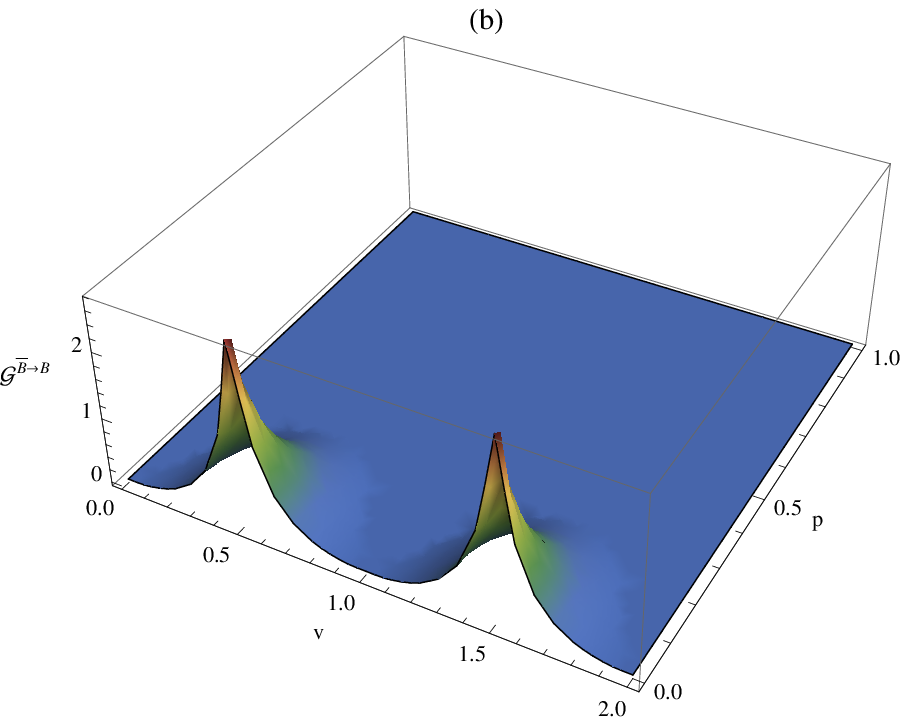}
\caption{ (color online). The Gaussian quantum steerability $ {\cal G}^{B \to \bar B}$ (left) and $ {\cal G}^{\bar B \to B}$ (right)  as functions of   the curvature parameter $p$ and mass parameter $\nu$ of the de Sitter space. The initial  squeezing parameter is fixed as $s=0.5$.}\label{Fig3}
\end{figure}

In Fig. (\ref{Fig3}) we plot  the Gaussian quantum steerability  between Bob and anti-Bob  as functions of   $p$ and $\nu$  with fixed squeezing  $s=0.5$. It is shown that  quantum steerability are  generated between Bob and Anti-Bob when  the curvature parameter $p$ is very small and the mass parameter are around  $\nu=1/2$  and $\nu=3/2$ . We find that both  the $ {\cal G}^{B \to \bar B}$ and  $ {\cal G}^{\bar B \to B}$ steerability are generated   when the curvature becomes stronger and stronger.    That is to say, Bob and antiBob can steer each other  when the curvature is strong enough  even though they are separated by the event horizon,  which verifies the fact that the  quantum steerability is one kind of  nonlocal quantum correlation.

\begin{figure}[htbp]
\centering
\includegraphics[height=2.0in,width=2.9in]{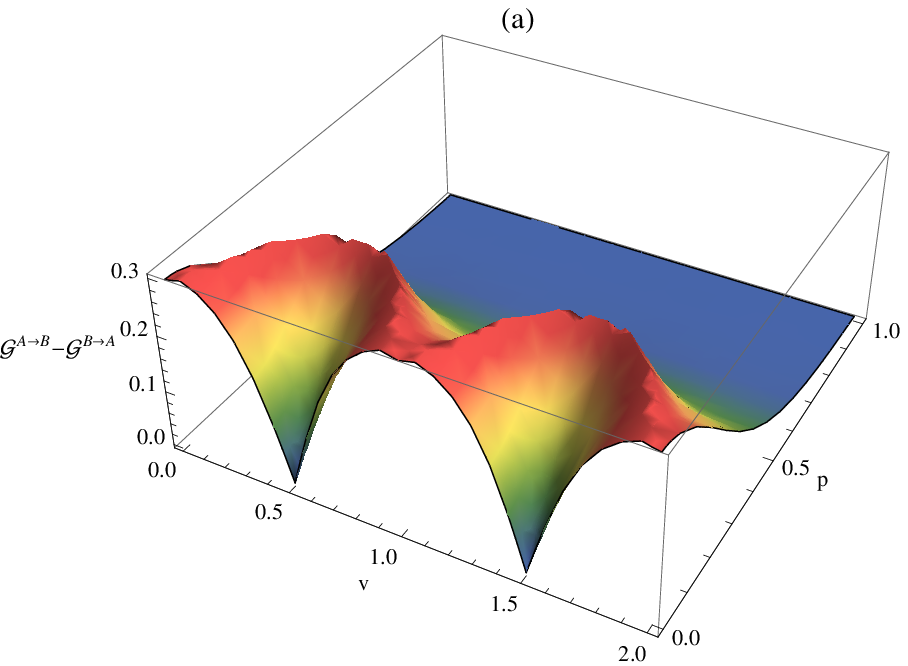}
\includegraphics[height=2.0in,width=2.9in]{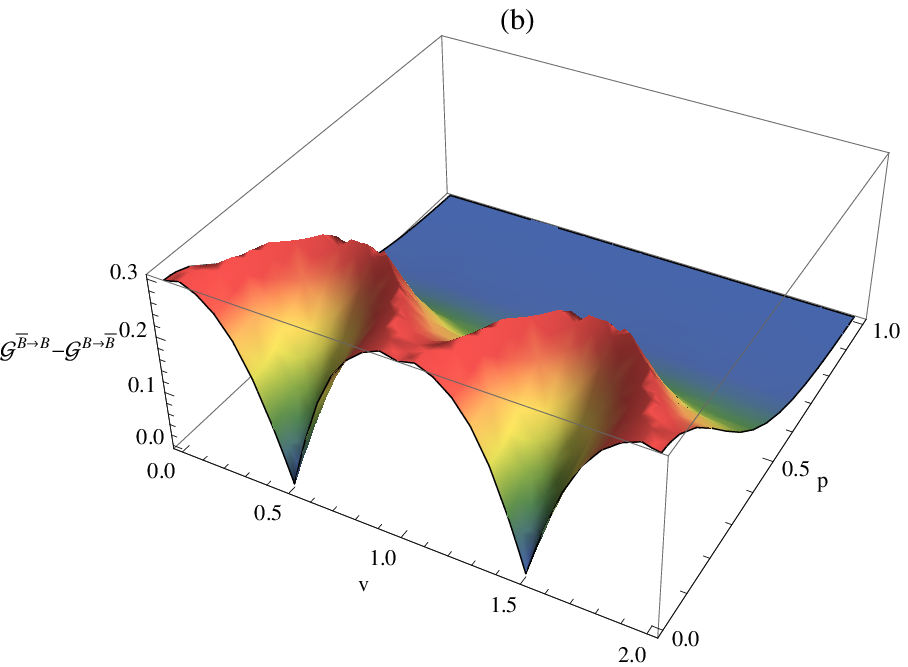}
\caption{ (color online). The Gaussian steerability asymmetry  between A and B  (left), as well as the   asymmetry  between B and $\bar B$ (right)  as functions of   the curvature parameter $p$ and mass parameter $\nu$ of the de Sitter space. The initial  squeezing parameter is fixed as $s=0.5$.}\label{Fig4}
\end{figure}

To check the degree of steerability asymmetric in de Sitter space,  we compute the Gaussian steerability asymmetry  $ |{\cal G}^{A \to B}-{\cal G}^{B \to A}|$ and $ |{\cal G}^{B \to \bar B}-{\cal G}^{\bar B \to B}|$.   In Fig. (4) we plot the Gaussian steerability asymmetry  between Alice and Bob, as well as the   asymmetry  between Bob and anti-Bob   as functions of   the curvature and mass parameters. As shown in Fig. (4a),
the steerability asymmetry between Alice and Bob increases with decreasing curvature parameter $p$, which demonstrates that the space curvature destroys the symmetry of initial  steerability. It is shown that  the condition which sets maximizing the steerability asymmetry  between Alice and Bob is $|\gamma_B|=\sqrt{\frac{1-sech(2s)}{2}}$, which  is exactly the ``sudden death" condition of   $A \to B$ steerability   in Fig. (2). That is,  the steerability asymmetry takes the maximum value when the state is un-steerable in the $A \to B$ direction.   The attaining of the peak of steerability  asymmetry indicates the system  experiences a  transformation from bidirectional steerability to unidirectional   steerability.

In Fig. (4b) we can see that the  maximum steerability asymmetry between Bob and anti-Bob is also attained  at $|\gamma_B|=\sqrt{\frac{1-sech(2s)}{2}}$, which  is identical with the condition of  the   $A-B$ steering asymmetry. Different from the  Alice-Bob asymmetry, the  attainment of maximal steerability  asymmetry between  Bob and anti-Bob  indicates the transition from unidirectional  steerability to bidirectional steerability in the de Sitter space. It is interesting to note that, like the   $A-B$ steerability asymmetry, the steerability asymmetry disappear   in the limit of infinite curvature  $p\to 0$ for   $\nu=1/2$ (conformal) and $\nu=3/2$ (massless).  However, the steerability asymmetry disappears  because both the $A\to B$  and $B\to A$  steerability equal to zero in this limit. Differently, the  $B-\bar B $ steerability  is  symmetry when both the   $B \to \bar B $  and   $\bar B \to B $ steerability take their maximum values. This phenomenon is nontrivial because the quantum steerability is usually  asymmetry  in the flat spacetime \cite{Adesso2015}.  We know that the quantum steering is asymmetry in the flat space because the parameters $\mu_A$ and $ \mu_B$ are usually different from each other.
This may indicates that the subsystem  $B$ and  $\bar B $ is symmetry in the  conformal scalar and massless scalar limit, which  deserves further study.

\section{Conclusions}

 We have studied the distribution  of steerability   among the mode $A(B)$ described by Alice (Bob) in the de Sitter region $R$, and the complimentary mode $\bar B$ described by a hypothetical observer anti-Bob in  the  causally disconnected region $L$.
 We first  derive the Bogoliubov transformation
between the Euclidean vacuum and the open chart vacua and then  obtain  a  phase-space description of quantum state evolution for continuous variables. We find that the quantum steerability is apparently affected by the  curvature of de Sitter space for  $\nu=1/2 $ (conformal) and $\nu=3/2$ (massless).  It is shown that  the $A \to B$ steerability  suffers from  ``sudden death", which is quite different from the behaviors of   entanglement and discord because the latter always survives while  the former vanishes only in the limit of infinite curvature  \cite{Kanno16}.
In de Sitter space, Bob and antiBob can steer each other  when the curvature is strong enough  even though they are separated by the event horizon. To verify the asymmetric property of steerability in de Sitter space, we compare the  $A-B$ and $B-\bar B $  steerability asymmetry.  In addition,  the maximum  asymmetry are obtained  when the $A \to B$ steerability experiences ``sudden death".  That is to say, the attainment of maximal steerability asymmetry indicates a
transition between unidirectional  steerable and bidirectional steerable in de Sitter space. Unlike the flat space, the asymmetry of quantum steerability can be  completely  destroyed in the limit of infinite curvature for some special scalar fields.

\begin{acknowledgments}
This work is supported by the Science and Technology Planning Project of Hunan Province under Grant No. 2018RS3061; and  the  Natural Science Fund  of Hunan Province  under Grant No. 2018JJ1016; and the  National Natural Science Foundation
of China under Grant  No. 11675052 and No. 11475061.	

\end{acknowledgments}
\appendix
\onecolumngrid

\section{Covariance Matrix of a Two-Mode Squeezed State in phase space}
In this appendix, we compute  the covariance matrix of a two-mode
squeezed quantum state in phase space. 
Firstly we  introduce the
quantities $\hat{x}_{\bm i}$ and $\hat{p}_{\bm i}$ such that
\begin{eqnarray}
\label{eq:defq}
\hat{x}_{\bm i}^{B(\bar B)} &=& \frac{1}{\sqrt{2}}\left(\hat{a}_{\bm i}^{B(\bar B)}
+\hat{a}_{\bm i}^{{B(\bar B)}\dagger}\right)\\
\label{eq:defpi}
\hat{p}_{\bm i}^{B(\bar B)} &=& -\frac{i}{\sqrt{2}}\left(\hat{a}_{\bm i}^{B(\bar B)}
-\hat{a}_{\bm i}^{{B(\bar B)}\dagger}\right).
\end{eqnarray}

\par

Let us now introduce the characteristic function. It is a real function
defined on a four-dimensional real space, given by
\begin{equation}
\chi(\xi)=Tr\left[\hat{\rho} \hat{{\cal W}}(\xi)\right],
\label{eq:chi:def}
\end{equation}
where $\hat{\rho}$ is
obviously the density operator and  $\hat{{\cal W}}(\xi)$ is the Weyl operator, namely
$
\hat{{\cal W}}(\xi)={\rm e}^{i\xi ^{\rm T}\hat{R}}$,
with $\hat{R}=\left(\hat{x}_{\bm i}^B,\hat{p}^B_{\bm
    i},\hat{x}^{\bar B}_{-{\bm i}},\hat{p}^{\bar B}_{-{\bm
      i}}\right)^{\rm T}\equiv
\left(\hat{R}_1,\hat{R}_2,\hat{R}_3,\hat{R}_4\right)^{\rm T}$.
\par
As a Gaussian state,
the two-mode squeezed
state has a Gaussian characteristic function given by
$
\chi (\xi)
={\rm e}^{-\xi^{\rm T}\gamma \xi/4},
$
where $\gamma $ is the covariance matrix, related to the two-point
correlation functions by $\langle
\hat{R}_j\hat{R}_k\rangle=\gamma_{jk}/2+iJ_{jk}/2$. Here, $J$ is the
commutator matrix, $iJ_{j,k}=[\hat{R}_j,\hat{R}_k]$, given by $J=J_1
\oplus J_1$ with
$
J_1=
\begin{pmatrix}
0 & 1 \\
-1 & 0
\end{pmatrix}.
$
For a two-mode Gaussian state, the
characteristic function $\chi (\xi)$ (where the components of $\xi$
are denoted as $\xi_1$, $\xi_2$, $\xi_3$ and $\xi_4$) can be written as \cite{Martin16} 
 
  \begin{eqnarray}
\chi(\xi) &=& Tr\left[
 \hat{S}(\gamma_p)\vert 0^A_{\bm i},0^B_{-{\bm i}}\rangle \langle 
0^A_{\bm i},0^B_{-{\bm i}}\vert  \hat{S}^{\dagger}(\gamma_p)
{\rm e}^{i\xi_1\hat{R}_1+i\xi_2\hat{R}_2+i\xi_3\hat{R}_3+i\xi_4\hat{R}_4}\right] \\
&=& Tr\left[\vert 0^A_{\bm i},0^B_{-{\bm i}}\rangle \langle 
0^A_{\bm i},0^B_{-{\bm i}}\vert  \hat{S}^{\dagger}(\gamma_p)
{\rm e}^{i\xi_1\hat{R}_1+i\xi_2\hat{R}_2+i\xi_3\hat{R}_3+i\xi_4\hat{R}_4}
 \hat{S}(\gamma_p)\right] \\
&=& \sum_{n=0}^{\infty}\sum_{n'=0}^{\infty}
\langle n_{\bm i},n'_{-{\bm i}}
\vert 0^A_{\bm i},0^B_{-{\bm i}}\rangle \langle 
0^A_{\bm i},0^B_{-{\bm i}}\vert  \hat{S}^{\dagger}(\gamma_p)
{\rm e}^{i\xi_1\hat{R}_1+i\xi_2\hat{R}_2+i\xi_3\hat{R}_3+i\xi_4\hat{R}_4}
 \hat{S}(\gamma_p)\vert n_{\bm i},n'_{-{\bm i}}\rangle \\
&=& \langle 
0^A_{\bm i},0^B_{-{\bm i}}\vert  \hat{S}^{\dagger}(\gamma_p)
{\rm e}^{i\xi_1\hat{R}_1+i\xi_2\hat{R}_2+i\xi_3\hat{R}_3+i\xi_4\hat{R}_4}
 \hat{S}(\gamma_p)\hat{R}(\theta _k)\vert 0^A_{\bm i},0^B_{-{\bm i}}\rangle
\\
&=& \langle 
0^A_{\bm i},0^B_{-{\bm i}}\vert  \hat{S}^{\dagger}(\gamma_p)
{\rm e}^{i\xi_1\hat{R}_1+i\xi_2\hat{R}_2}{\rm e}^{+i\xi_3\hat{R}_3+i\xi_4\hat{R}_4}
 \hat{S}(\gamma_p)\vert 0^A_{\bm i},0^B_{-{\bm i}}\rangle \\
&=& {\rm e}^{i\xi_1\xi_2/2+i\xi_3\xi_4/2}\langle 
0^A_{\bm i},0^B_{-{\bm i}}\vert  \hat{S}^{\dagger}(\gamma_p)
{\rm e}^{i\xi_1\hat{R}_1}{\rm e}^{i\xi_2\hat{R}_2}{\rm e}^{i\xi_3\hat{R}_3}
{\rm e}^{i\xi_4\hat{R}_4}
 \hat{S}(\gamma_p)\vert 0^A_{\bm i},0^B_{-{\bm i}}\rangle,
\end{eqnarray}
where the Baker-Campbell-Hausdorff formula ${\rm
  e}^{\hat{A}+\hat{B}} ={\rm
  e}^{-[\hat{A},\hat{B}]/2}{\rm e}^{\hat{A}}{\rm
  e}^{\hat{B}}$ have been used, which is valid if the operators $\hat{A}$ and $\hat{B}$ commute with  $[\hat{A},\hat{B}]$. The next step is  introducing the operator $\hat{S}\hat{R}(\hat{S}\hat{R})^{\dagger}$
 between the  exponential factors. To this end  we
calculate  \cite{Martin16}
\begin{eqnarray}
 \hat{S}^{\dagger}(\gamma_p)
{\rm e}^{i\xi_1\hat{R}_1} \hat{S}(\gamma_p)
&=&  \hat{S}^{\dagger}(\gamma_p)
\sum_{n=0}^{\infty}\frac{1}{n!}i^n\xi_1^n\hat{R}_1^n
 \hat{S}(\gamma_p)\\
&=&
\sum_{n=0}^{\infty}\frac{1}{n!}i^n\xi_1^n
 \hat{S}^{\dagger}(\gamma_p)
\hat{R}_1^n
 \hat{S}(\gamma_p).
\end{eqnarray}

By using the fact that 
\begin{equation}
\hat{S}^{\dagger}\hat{R}_1^n\hat{S}
=\hat{S}^{\dagger}
\hat{R}_1\hat{S}\hat{S}^{\dagger}
\hat{R}_1^{n-1}\hat{S}
=\hat{S}^{\dagger}
\hat{R}_1\hat{S}\hat{S}^{\dagger}
\hat{R}_1\hat{S}\hat{S}^{\dagger}
\hat{R}_1^{n-2}\hat{S}
=\left(\hat{S}^{\dagger}\hat{R}_1\hat{S}\right)^n,
\end{equation}
 one can rewrite the series as  \cite{Martin16}
\begin{eqnarray}
\hat{S}^{\dagger}(\gamma_p)
{\rm e}^{i\xi_1\hat{R}_1}\hat{S}(\gamma_p)
&=&
\sum_{n=0}^{\infty}\frac{1}{n!}i^n\xi_1^n
\left[\hat{S}^{\dagger}(\gamma_p)
\hat{R}_1
\hat{S}(\gamma_p)\right]^n\\
&=& {\rm e}^{i\xi_1
\hat{S}^{\dagger}(\gamma_p)\hat{R}_1
\hat{S}(\gamma_p)}.
\end{eqnarray}
Using this result  for the four terms of the characteristic function, we
arrive at
\begin{eqnarray}
\label{eq:chiinter}
\chi (\xi) &=& 
{\rm e}^{i\xi_1\xi_2/2+i\xi_3\xi_4/2}\langle 
0^A_{\bm i},0^B_{-{\bm i}}\vert 
{\rm e}^{i\xi_1\hat{S}^{\dagger}\hat{R}_1\hat{S}}
{\rm e}^{i\xi_2\hat{S}^{\dagger}\hat{R}_2\hat{S}}
{\rm e}^{i\xi_3\hat{S}^{\dagger}\hat{R}_3\hat{S}}
{\rm e}^{i\xi_4\hat{S}^{\dagger}\hat{R}_4\hat{S}}
\vert 0^A_{\bm i},0^B_{-{\bm i}}\rangle.
\end{eqnarray}
  
To proceed, one has to evaluate the four terms
$\hat{S}^{\dagger}\hat{R}_i\hat{S}$, which are  \cite{Martin16}
\begin{eqnarray}
\hat{\Omega}_1 &\equiv& \hat{S}^{\dagger}(\gamma_p)
\hat{R}_1\hat{S}(\gamma_p) = (\hat{R}_1-\hat{R}_2) \cosh \gamma_p
+(\hat{R}_3+\hat{R}_4) \sinh \gamma_p,\\
\hat{\Omega}_2 &\equiv&\hat{S}^{\dagger}(\gamma_p)
\hat{R}_2
\hat{S}(\gamma_p) = (\hat{R}_1+\hat{R}_2)\cosh \gamma_p
+ (\hat{R}_3-\hat{R}_4) \sinh \gamma_p,\\
\hat{\Omega}_3 &\equiv&
\hat{S}^{\dagger}(\gamma_p)\hat{R}_3
\hat{S}(\gamma_p) =  (\hat{R}_1+\hat{R}_2) \sinh \gamma_p
+ (\hat{R}_3-\hat{R}_4) \cosh \gamma_p,\\
\hat{\Omega }_4 &\equiv&
\hat{S}^{\dagger}(\gamma_p)\hat{R}_4
\hat{S}(\gamma_p) =  (\hat{R}_1-\hat{R}_2) \sinh \gamma_p
+(\hat{R}_3+\hat{R}_4) \cosh \gamma_p.
\end{eqnarray}
Then we find that  the characteristic function
takes the form  \cite{Martin16}
\begin{eqnarray}
\chi(\xi) &=&
{\rm e}^{i\xi_1\xi_2/2+i\xi_3\xi_4/2}\langle
0^B_{\bm i},0^{\bar B}_{-{\bm i}}\vert
{\rm e}^{-i\xi_1\xi_2/2}
{\rm e}^{i\xi_1\hat{\Omega}_1+i\xi_2\hat{\Omega}_2}
{\rm e}^{-i\xi_3\xi_4/2}
{\rm e}^{i\xi_3\hat{\Omega}_3+i\xi_4\hat{\Omega} _4}
\vert 0^B_{\bm i},0^{\bar B}_{-{\bm i}}\rangle \\
&=&
\langle 0^B_{\bm i},0^{\bar B}_{-{\bm i}}\vert
{\rm e}^{i\xi_1\hat{\Omega}_1+i\xi_2\hat{\Omega}_2+i\xi_3\hat{\Omega}_3+i\xi_4\hat{\Omega }_4}
\vert 0^B_{\bm i},0^{\bar B}_{-{\bm i}}\rangle
 \\
&=&
\label{eq:chiinter2}
\langle 0^B_{\bm i},0^{\bar B}_{-{\bm i}}\vert
{\rm e}^{i\eta_1\hat{R}_1+i\eta_2\hat{R}_2+i\eta_3\hat{R}_3+i\eta_4\hat{R}_4}
\vert 0^B_{\bm i},0^{\bar B}_{-{\bm i}}\rangle \\
& \equiv&
\chi_{\rm vac}(\eta_1,\eta_2,\eta_3,\eta_4),
\end{eqnarray}
where the coefficients $\eta_i$ can be expressed as
\begin{eqnarray}
\eta_1 &=& \xi_1 \cosh \gamma_p  +(\xi_2+\xi_3+\xi_4) \sinh \gamma_p, \\
\eta_2 &=&( -\xi_1 +\xi_2)\cosh \gamma_p
+(\xi_3-\xi_4) \sinh \gamma_p, \\
\eta_3 &=& (\xi_1+\xi_2) \sinh \gamma_p
+(\xi_3 +\xi_4)\cosh \gamma_p,\\
\eta_4 &=& (\xi_1-\xi_2) \sinh \gamma_p
+(-\xi_3+\xi_4) \cosh \gamma_p.
\end{eqnarray}
Since
the definition of the covariance matrix $\gamma$ is given by the
expression  \cite{Martin16}
\begin{equation}
\label{eq:chiGauss}
\chi (\xi_1,\xi_2,\xi_3,\xi_4)
={\rm e}^{-\xi^{\rm T}\gamma \xi/4}.
\end{equation}
It  implies that
\begin{equation}
\sum_{i=1}^{4}\eta_i^2=\sum_{i=1}^4\sum_{j=1}^4\gamma_{ij}\xi_i\xi_j,
\end{equation}
which allows us to infer the components of the covariance
matrix.

Using the above expressions of the correlation matrixes, the explicit
form of the two-point correlators are obtained by some lengthy but straightforward calculations, which are
\begin{eqnarray}
\label{eq:qq}
\langle \hat{x}^B_{\bm i}\hat{x}^{B}_{\bm i}\rangle &=&
\langle \hat{x}^{\bar B}_{-{\bm i}}\hat{x}^{\bar B}_{-{\bm i}}\rangle =
\cosh(2\gamma_p),
\quad
\langle \hat{p}^B_{\bm i}\hat{p}^B_{\bm i}\rangle =
\langle \hat{p}^{\bar B}_{-{\bm i}}\hat{p}^{\bar B}_{-{\bm i}}\rangle
=cosh(2\gamma_p)
\\
\label{eq:qminusq}
\langle \hat{x}^B_{\bm i}\hat{x}^{\bar B}_{-{\bm i}}\rangle &=&
\frac{1}{2}\sinh(2\gamma_p),
\quad
\langle \hat{p}^B_{\bm i}\hat{p}^{\bar B}_{-{\bm i}}\rangle
=
-\frac{1}{2}\sinh(2\gamma_p),
\\
\label{eq:qminuspi}
\langle \hat{x}^B_{\bm i}\hat{p}^{\bar B}_{-{\bm i}}\rangle &=&
\langle \hat{p}^B_{\bm i}\hat{x}^{\bar B}_{-{\bm i}}\rangle
=
\frac{1}{2}\sinh(2\gamma_p),
\quad.
\end{eqnarray}
Then one obtains the
following covariance matrix 
\begin{equation}
\sigma_{B\bar B}(\gamma_B)=
\begin{pmatrix}
\cosh \left(2\gamma_p\right) & 0 & \sinh \left(2\gamma_p\right)
& \sinh\left( 2\gamma_p\right)
 \\
0 & \cosh \left(2\gamma_p\right) & \sinh \left(2\gamma_p\right)
& -\sinh\left( 2\gamma_p\right)  \\
\sinh \left(2\gamma_p\right)
& \sinh \left(2\gamma_p\right)
& \cosh\left(2\gamma_p\right) & 0 \\
\sinh\left( 2\gamma_p\right)
& -\sinh\left( 2\gamma_p\right)
& 0 & \cosh\left(2\gamma_p\right)
\end{pmatrix},
\end{equation}
where $\cosh \gamma_B=(\sqrt{1-|\gamma_B|^2})^{-1}$.

\section{The  definition and measurement  of  quantum steerability}
In this appendix we introduce the  conception and definition of  quantum steerability. 
Under a set of measurements $\mathcal{M}_A$ on Alice, a bipartite system is $A\to B$ steerable---i.e., Alice can steer Bob---\textit{iff} it is \textit{not} possible
for every pair of local observables $R_A \in \mathcal{M}_A$ on $A$ and $R_B$  on $B$ (with respective outcomes $r_A$ and $r_B$), to express the joint probability as   \cite{wiseman}
\begin{equation}\label{C1}
P\left( {{r_A},{r_B}|{R_A},{R_B},{\rho _{AB}}} \right) = \sum\limits_\lambda  {{\wp_\lambda }} \, \wp\left( {{r_A}|{R_A},\lambda } \right)P\left( {{r_B}|{R_B},{\rho _\lambda }} \right),
\end{equation}
where  ${\wp_\lambda }$ and $\wp \left( {{r_A}|{R_A},\lambda }\right)$  are
probability distributions, involving
the local hidden variable $\lambda$. In addition,  $P\left( {{r_B}|{R_B},{\rho _\lambda }} \right)$  is the conditional  probability distribution associated to the extra condition of being evaluated on the state $\rho_\lambda$.
That is, at least one measurement pair $R_A$ and $R_B$ must violate the expression in Eq. (C1)  when ${\wp_\lambda }$ is fixed across all measurements. The probability distribution  $P\left( {{r_B}|{R_B},{\rho _\lambda }} \right)$  means that a complete knowledge of Bob's devices (but not of Alice's ones) is required to formulate the steering condition  \cite{wiseman}.

Here we consider  the fully Gaussian scenario, where the initial state is a Gaussian state and the  observers' measurement sets $\mathcal{M}_{A,B}$ are also Gaussian (i.e., mapping Gaussian states into Gaussian states). A Gaussian measurement can be described by a positive Gaussian operator with covariance matrix $T^{{R_A}}$, satisfying  \begin{equation}{T^{{R_A}}} + i\,{\Omega _A} \ge 0. \end{equation}  Once Alice makes a measurement $R_A$ and gets an outcome $r_A$, Bob's conditioned state $\rho_{B}^{r_A|R_A}$ is Gaussian.  The covariance matrix of Bob after Alice's measurement is   given by   \cite{wiseman} \begin{equation}B_{}^{{R_A}} = B - C{\left( {{T^{{R_A}}} + A} \right)^{ - 1}}{C^{\sf T}}, \end{equation} which is independent of Alice's outcome.

As has been shown in \cite{wiseman}, a Gaussian state $\rho_{AB}$ is $A\to B$ steerable by Alice's Gaussian measurements \textit{iff} the condition
\begin{equation}\label{nonsteer}
{\sigma _{AB}} + i\,({0_A} \oplus {\Omega _B}) \ge 0,
\end{equation}
is violated. Henceforth, a violation of Eq. \eqref{nonsteer} is necessary and sufficient for Gaussian  $A\to B$ steerability.
Writing this in matrix form, using the covariance matrix in  Eq. (\ref{CM}) of the main manuscript, the nonsteerability inequality Eq. (\ref{nonsteer}) is equivalent to two simultaneous conditions: (i) $A > 0$, and (ii) ${M^B_{\sigma}} + i{\Omega _B} \ge 0$, where $M^B_{\sigma} = B - {C^{\sf T}}{A^{ - 1}}C$ is the Schur complement of $A$ in the covariance matrix  $\sigma_{AB}$  \cite{wiseman}.
Condition (i) is always verified for  any  physical covariance matrix. Therefore, $\sigma_{AB}$ is $A \to B$ steerable \textit{iff} the symmetric  matrix $M^B_{\sigma}$ is not a {\it bona fide} covariance matrix, i.e., if condition (ii) is violated \cite{wiseman}.
According to Williamson's theorem \cite{williamson}, the symmetric matrix $M^B_{\sigma}$ is diagonalized by a symplectic transformation $S_B$  such that $S_B M^B_{\sigma} S_B^{\sf T}=\text{diag}\{\bar{\nu}^B_1,\bar {\nu}^B_1,\ldots,\bar{\nu}^B_m,\bar {\nu}^B_m\}$, where $\{\bar{\nu}^B_{j}\}$ are the symplectic eigenvalues of $M^B_{\sigma}$.
Then  the degree of  $A \to B$ steerability can be measured  by  \cite{Adesso2015}
\begin{equation}\label{GSAB}
{\cal G}^{A \to B}(\sigma_{AB}):=
\max\bigg\{0,\,-\sum_{j:\bar{\nu}^B_j<1} \ln(\bar{\nu}^B_j)\bigg\}\,,
\end{equation}
which quantifies the amount by which the condition given by Eq. (\ref{nonsteer}) fails to be fulfilled.
This is the  Gaussian $A \to B$ steerability, which is invariant under local symplectic operations, it vanishes \textit{iff} the state described by  Eq. (\ref{CM}) of the main manuscript is nonsteerable by Gaussian measurements. In other words, the  $A \to B$ steerability in fact quantifies the degree by which the condition (\ref{nonsteer}) fails to be fulfilled by Alice's measurement. 

Clearly, the Gaussian $B \to A$ steering can also be obtained by swapping the roles of $A$ and $B$, resulting in an expression like Eq. (\ref{GSAB}), in which the symplectic eigenvalues of the  Schur complement of $B$,   $M^A_{\sigma} =  A - {C}{B^{ - 1}}{C^{\sf T}}$,  appear instead.  When the steered party has one mode only,  $M^B_{\sigma}$ has a single symplectic eigenvalue  \begin{equation}\bar{\nu}^B = \sqrt{\det M^B_{\sigma}}. \end{equation} By defining the  Schur complement $\det{\sigma_{AB}} = \det{A} \det{M^B_{\sigma}}$, one can obtain that 
\begin{equation}
{\cal G}^{A \to B}(\sigma_{AB}) =
\mbox{$\max\big\{0,\, \frac12 \ln {\frac{\det A}{\det \sigma_{AB}}}\big\}$}= \max\big\{0,\, {\cal S}(A) - {\cal S}(\sigma_{AB})\big\}\,, \label{GS1}
\end{equation}
which is Eq. (14) of the main manuscript. To compute the quantum steering we only need to 
get the  interplay between the global purity $\mu = 1/\sqrt{\det \sigma_{AB}}$ and the one-side purities $\mu_{A(B)} = 1/\sqrt{\det A (B)}$. Employing the ratio $\eta = (\mu_A \mu_B)/\mu$, one finds that all physical two-mode Gaussian states live in the region $\eta_0 \leq \eta \leq 1$ where $\eta_0=\mu_A \mu_B + |\mu_A - \mu_B|$ \cite{Adesso2015}. States with $\eta_s \leq \eta \leq 1$ where $\eta_s=\mu_A + \mu_B - \mu_A \mu_B $ are necessarily separable,  states with $\eta_0 \leq \eta < \eta_e$ are  entangled \cite{Adesso2015}, where $\eta_e=\sqrt{\mu_A^2+\mu_B^2-\mu_A^2\mu_B^2}$. Within the entangled region, states which satisfy  $\eta \geq \{\mu_A, \mu_B\}$ are nonsteerable; states with $\eta < \mu_B$ are $A \to B$ steerable; states with $\eta < \mu_A$ are $B \to A$ steerable. For this reason  we can see that the quantum steering is asymmetry in the flat space because $\mu_A$ and $ \mu_B$ are usually different from each other. In fact, 
the asymmetry of steering for  Gaussian states  in flat space has been experimentally demonstrated in \cite{Handchen}. 

\section{Covariance matrix for the final state of the entire system}

The final state given in  Eq. (19) of  the main manuscript is the phase space  description of the tripartite system after the curvature-induced  squeezing transformation. The covariance matrix  for the final state of the entire system is \cite{wang20}
\begin{eqnarray}
\nonumber\sigma^{\rm }_{AB \bar B}(s,r) &=& \big[I_A \oplus  S_{B,\bar B}(\gamma_p)\big] \big[\sigma^{\rm (M)}_{AB}(s) \oplus I_{\bar B}\big]\\&& \nonumber\big[I_A \oplus  S_{B,\bar B}(\gamma_p)\big]\\
 &=& \left(
       \begin{array}{ccc}
          \mathcal{\sigma}_{A} & \mathcal{E}_{AB} & \mathcal{E}_{A\bar B} \\
         \mathcal{E}^{\sf T}_{AB} &  \mathcal{\sigma}_{B} & \mathcal{E}_{B\bar B} \\
         \mathcal{E}^{\sf T}_{A\bar B} & \mathcal{E}^{\sf T}_{B\bar B} &  \mathcal{\sigma}_{\bar B} \\
       \end{array}
     \right)
 \,,
\end{eqnarray}
where $\sigma^{\rm (G)}_{AB}(s) \oplus I_{\bar B}$ is the initial state of  the entire system.
In Eq. (C1) the diagonal elements have the following forms: $ \mathcal{\sigma}_{A}=\cosh(2s)I_2$, $\mathcal{\sigma}_{B}=[\cosh(2s) \cosh^2(\gamma_p) + \sinh^2(\gamma_p)]I_2$, $\mathcal{\sigma}_{\bar B}=[\cosh^2(\gamma_p) + \cosh(2s) \sinh^2(\gamma_p)]I_2$.  The non-diagonal elements are 
$\mathcal{E}_{AB}=[\cosh(\gamma_p) \sinh(2s)]Z_2$, $\mathcal{E}_{B\bar B}=[\cosh^2(s) \sinh(2\gamma_p)]Z_2$,and $\mathcal{E}_{A\bar B}=[\sinh(2s) \sinh(\gamma_p)]Z_2$,  where
\begin{equation}
\label{eq:def1}
Z_2=
\begin{pmatrix}
1 & 0 \\
0 & -1
\end{pmatrix}.
\end{equation}

\end{document}